\documentclass[letterpaper]{article}
\usepackage[a4paper]{geometry}
    \usepackage{setspace}
\usepackage{changepage} %% so that if you have a table, you can make it not go off the page! YAY
\usepackage[american]{babel}
\usepackage{graphicx}
\usepackage{blindtext}
\usepackage{amssymb}% http://ctan.org/pkg/amssymb
\usepackage{amsmath}
\usepackage{mathrsfs}
\usepackage{pifont}% http://ctan.org/pkg/pifont
\usepackage{epigraph}
\usepackage{setspace}

\usepackage{hyperref}
\hypersetup{
colorlinks=true,
    linkcolor=blue,
    filecolor=magenta,      
    urlcolor=blue,
    citecolor=blue,
    pdfpagemode=FullScreen,
    }

\usepackage[authoryear]{natbib}

\title{On Gauge Gravity's\\`Hypermomentum Challenge'\\ to the Geometric Trinity}
\author{Kartik Tiwari\thanks{Lichtenberg Group for History and Philosophy of Physics, Institute of Philosophy, University of Bonn; Argelander Institute for Astrophysics, University of Bonn}}

\begin{document}
\maketitle

\begin{abstract}
Claims concerning a dynamical equivalence between general relativity and two reformulations of it, in which gravity is encoded in torsion and non-metricity respectively, have attracted some attention. If taken to be true, this `Geometric Trinity' suggests that it is impossible to determine, even in principle, whether one inhabits a spacetime solution possessing curvature, torsion or non-metricity (insofar as the probes of geometry are paths of test particles). This claim is in direct tension with the long tradition of studying gravity as a gauge theory, in which distinct geometric attributes couple to distinct properties of matter (such as energy-momentum to curvature and spin angular momentum to torsion). In this short paper, I compare these two conflicting frameworks for geometrizing spacetime and describe the precise source, nature and implications of their disagreement. The equivalence claim holds only under a substantive restriction on the matter sector, namely that matter does not couple to an independent affine connection. Because the relevant couplings remain beyond current empirical reach, the choice between the frameworks turns on a trade-off among ontological, epistemic, and methodological forms of parsimony.
\end{abstract}
\singlespacing

\tableofcontents
\epigraph{What is the use then of imagining an electro-tonic state of which we have no distinctly physical conception, instead of a formula of attraction which we can readily understand? I would answer, that it is a good thing to have two ways of looking at a subject, and to admit that there are two ways of looking at it.}{- James Clerk Maxwell,\\ \textit{On Faraday's Lines of Force }(1855–56)}

\section{Introduction}

The geometrization of spacetime in any physical theory is contingent on several choices that, in turn, invite vivid disagreements. While some such disagreements reveal their forms transparently, others take place between theories that superficially resemble each other, say, by utilizing similar formal structures. In this short paper, I compare two conflicting frameworks for geometrizing spacetime that use the same mathematical objects (spacetimes possessing geometric attributes besides curvature) but make claims that are in tension with each other. In doing so, I describe the precise source, nature and implication of this disagreement and, finally, provide an account of the epistemic virtues available in defense of each framework. 

I begin by briefly introducing the key geometrical object (metric-affine spacetimes) and the properties it admits (curvature, torsion and non-metricity). Then, I introduce the two frameworks in tension: the `Geometric Trinity' (GT) framework \citep{jimenez_geometrical_2019} and the gauge gravity framework, represented here by `Metric Affine Gauge Gravity' (MAG) \citep{hehl_metric-affine_1995}. Starting with the former, I summarize the argument at the core of the Geometric Trinity discourse and some of the ways these formal results have been interpreted (in particular, concerning their bearing on empirical distinguishability, under-determination and geometric conventionalism). In contrasting the gauge gravity framework, I lead with the sketch of a simpler theory (Einstein-Cartan gravity) to show how the incompatibility with Geometric Trinity arises already and generalize this tension to the more interesting case of Metric Affine Gravity. While the Geometric Trinity claims a formal equivalence between gravitational theories based on curvature, torsion and non-metricity respectively, the gauge gravity framework directly denies any such equivalence. 

This incompatibility is sourced from what I call the `hypermomentum challenge', whose technical content was first made explicit by \citet{iosifidis_motion_2024}, though the underlying tension has been latent in gauge gravity literature since \citet{hehl_metric-affine_1995}. In \S \ref{sec:epv}, I anticipate some objections and counter-objections to the hypermomentum challenge available in support of Geometric Trinity and Gauge Gravity respectively. 

Despite operating within overlapping theoretical territory, the two research communities have developed largely distinct bodies of literature, with rare direct engagement across the divide. Each brings its own notational conventions and community-specific vocabulary. One aim of this paper is to remain accessible to researchers on both sides of this disagreement and where exposition may seem familiar to one audience, it is intended as orientation for the other.

\subsection{Mathematical Preliminaries}\label{prelims}
I begin by defining the geometric objects that are referenced frequently through this paper. The most commonly discussed tensorial object relevant to this work is the Riemann tensor that represents the curvature of a spacetime manifold, which can be defined for a general affine connection $\Gamma$ as follows
\begin{equation}
    R^{\alpha}{}_{\beta\mu\nu}(\Gamma) \equiv \partial_{\mu} \Gamma^{\alpha}_{\nu\beta} - \partial_{\nu} \Gamma^{\alpha}_{\mu\beta} + \Gamma^{\alpha}_{\mu\lambda} \Gamma^{\lambda}_{\nu\beta} - \Gamma^{\alpha}_{\nu\lambda} \Gamma^{\lambda}_{\mu\beta}.
    \label{eq:placeholder_label}
\end{equation}
Curvature captures a vector's deviation from itself on parallel transporting over a closed loop and is typically the geometric quantity understood as being due to the presence of energy-momentum in spacetime. The next useful object to define is the covariant derivative of the metric itself
\begin{equation}
    Q_{\alpha\mu\nu} \equiv \nabla_{\alpha} g_{\mu\nu},
    \label{eq:placeholder_label}
\end{equation}
which is called non-metricity and it measures the change in a vector's length as it is parallel transported over a manifold. Now, consider a manifold which admits an affine connection such that the bottom two indices do not commute (as opposed to the usual case of Levi-Civita connection used in general relativity). I denote this quantity, in a holonomic basis, as
\begin{equation}
    T^{\alpha}{}_{\beta\gamma} \equiv \Gamma^\alpha_{\beta\gamma} - \Gamma^\alpha_{\gamma\beta}
\end{equation}
where $T$ is a bilinear map associated with the affine connection and is called the Cartan torsion tensor (or, simply, torsion). The torsion tensor $T(X,Y)$ takes vectors $X$ and $Y$ as its input and maps them to an output vector $T(X, Y)$ representing the displacement in the tangent space when the tangent space is rolled along an infinitesimal parallelogram with sides $X$ and $Y$. Similar to how curvature is heuristically interpreted as a vector's deviation from itself when parallel transported along a closed loop, the torsion tensor can be interpreted as the non-closure of a parallelogram in that manifold. 

In General Relativity, one restricts attention to spacetime geometries in which the straightest lines are also the distance extremizing ones \citep{levi-civita_nozione_1917, wald_general_1984}. The notion of straightness is endowed by the affine connection $\Gamma^\alpha_{\beta\gamma}$ and the concept of distance comes from the metric $g_{\alpha\beta}$. Roughly speaking, the ‘straightness$\iff$distance-extremizing’ condition can be understood as imposing two separate mathematical restrictions:
\begin{itemize}
    \item The affine connection is symmetric in its bottom two indices i.e. $\Gamma^\alpha_{\beta\gamma} - \Gamma^\alpha_{\gamma\beta} = 0 \implies T^\alpha_{\beta\gamma}=0$. This condition imposes vanishing spacetime torsion.
    \item The non-metricity tensor $Q$ is zero i.e. $\nabla_{\alpha} g_{\mu\nu} = 0 \implies Q_{\alpha\mu\nu} = 0$. This is often called the `metric-compatibility' condition.
\end{itemize}
Thus, the spacetime manifolds that form the arena of general relativity are torsion-free and metric-compatible. 

In exploring alternate geometrical descriptions of gravity that still lie in a conceptual neighborhood of general relativity \citep{lehmkuhl_introduction_2017}, one might wish to suspend these assumptions and explore their consequences. The geometrical objects that violate these assumptions are called metric affine spacetime manifolds and they contain attributes, namely torsion and non-metricity, that are assumed to vanish in GR. The existing scholarship on gravitational physics with such alternatives presents two conflicting frameworks based on two distinct sets of theory-building methods and foundational assumptions. Crucially, the two frameworks employ the same geometrical objects for constructing their respective theories of gravity and both take as their starting point the observation that general relativity occupies only one region in a wider theory building space (as it uses only metric-compatible torsion-free manifolds). I now discuss each framework in more detail and then examine how they relate to each other. 

\section{The Geometric Trinity of Gravitational Theories}\label{sec:GT}
It has been known since the early days of GR that there exists a family of theories of gravity constructed from torsion rather than curvature, usually called teleparallel theories of gravity (TGR). Historically, torsion-gravity can be traced back to Einstein’s \emph{Fernparallelismus} programme, where distant parallelism was explored in pursuit of a unified field theory of gravitation and electromagnetism \citep{sauer_field_2004}. That unified-field project was distinct from TGR's later revival, in which teleparallelism attracted renewed interest because it was believed it could be cast as a gauge theory of spacetime translations \citep{cho_einstein_1976, cho_gauge_1976, maluf_teleparallel_2013} and because it appeared to offer a cleaner handle on questions such as the definition of gravitational energy-momentum \citep{moller_conservation_1961}. However, teleparallel theories have also attracted criticism from both philosophers and physicists. The analogy with Yang-Mills gauge theory is imperfect \citep{hehl_metric-affine_1995, weatherall__2025}, the formalism appears to involve surplus structure \citep{march_equivalence_2025, weatherall_are_2025, weatherall__2025}, and there are concerns regarding the mathematical coherence of teleparallel theories \citep{duerr_read_teleparallel_forthcoming}. 

More recently, it has been shown that one may formulate a similar family of theories using non-metricity rather than torsion; these are known as symmetric teleparallel theories of gravity (STGR) \citep{nester_symmetric_1999, adak_lagrange_2006}. Similar to TGR, STGR has been taken to possess several mathematically attractive features\footnote{In particular, the so-called `coincident gauge' can trivialize the affine connection \citep{jimenez_coincident_2018, jimenez_geometrical_2019, read_clarifying_2026}, while \(f(Q)\) extensions exhibit features of cosmological interest \citep{jimenez_cosmology_2020}.}. But STGR, too, inherits some of the same worries as TGR, including skepticism about surplus structure and about its gauge-theoretic interpretation.

For the purposes of this paper, I can largely ignore these debates and, instead, focus on one particular cluster of claims concerning TGR and STGR that, from hereon, I call the `Geometric Trinity' framework. The first important GT claim here is that there exists a theory belonging to the family of teleparallel theories, defined on manifolds with vanishing curvature and non-metricity, that is equivalent to General Relativity (under a restricted equivalence criterion stated precisely in the next subsection), but in which matter influences, and is influenced by, spacetime torsion rather than curvature. This special member of the TGR family is named the Teleparallel Equivalent of General Relativity (TEGR) \citep{hayashi_new_1979, jimenez_geometrical_2019}. Similarly, the second important GT claim is that there exists a theory belonging to the family of symmetric teleparallel theories, defined on manifolds with vanishing curvature and torsion, that is equivalent to General Relativity under the same equivalence criterion, but in which matter influences, and is influenced by, spacetime non-metricity rather than curvature. \textit{Mutatis mutandis}, this theory is called the Symmetric Teleparallel Equivalent of General Relativity (STEGR)\citep{nester_symmetric_1999, jimenez_coincident_2018, jimenez_geometrical_2019}. Thus, from the two aforementioned claims and the transitivity of the equivalence relation, I summarize the central proposal of Geometric Trinity as follows: there exists a trio of equivalent theories of gravity which are constructed exclusively using just one of the three possible, independent geometrical attributes of spacetime i.e. GR (using curvature), TEGR (using torsion) and STEGR (using non-metricity) \citep{jimenez_geometrical_2019, capozziello_comparing_2022}\footnote{The Newtonian limits of TEGR and STEGR have also been derived and shown to be mutually equivalent \citep{wolf_non-relativistic_2024}, providing a useful consistency check for the GT framework. One can already anticipate the contrasting framework, called Metric Affine Gauge Theory of Gravity (or just metric affine gravity or gauge gravity), in which a theory of gravity is built without restricting it to only one geometric object at a time. Instead, the gauge gravity framework treats spacetimes with varying degrees of all three geometrical attributes as valid solutions to field equations.}.  

To evaluate this proposal and identify its limitations, it is crucial to understand the precise sense in which the Geometric Trinity claims an equivalence between the three types of theories\footnote{For more general discussions of the criteria for equivalence between physical theories, see discussions in \citet{weatherall_theoretical_2018, north_equivalence_2021, weatherall_are_2025}}. Clearly, the three nodes of the trinity admit different geometric intuitions and interpretations (in that they seem to suggest different ontologies for how the world may be). The different geometric objects have distinct mathematical properties (some examples of why they might bear on familiar phenomena are presented in \S \ref{implications}) and the theories are not equivalent in a category theoretic sense either \citep{weatherall_are_2025}.   The equivalence criterion that is being discussed here, instead, refers to a `dynamical equivalence' relation i.e. the metric field equations that govern the dynamics in these theories coincide, for reasons I summarize below. In the remainder of this section, I sketch the argument from \citet{jimenez_geometrical_2019} for the dynamical equivalence within the Geometric Trinity and discuss its key implications. In the subsequent sections, I introduce the gauge gravity framework and describe the challenges it poses to this dynamical equivalence claim.

 \subsection{Reasons for Dynamical Equivalence}\label{sec:GT_Arg}
I begin with the well-known Einstein-Hilbert action defined over Riemannian spacetimes\footnote{When I use the term `Riemannian spacetimes', I do not refer to manifolds with positive-definite signature but, instead, follow the convention of gauge-gravity literature to refer to Lorentzian manifolds equipped with Levi-Civita connections.}, 
\begin{equation}
    S_{\mathrm{GR}_{(1)}} = \frac{1}{16\pi G} \int \mathrm{d}^4 x \sqrt{-g} \, \mathcal{R}(g)
    \label{eq:1},
\end{equation}
where $ \mathcal{R}(g)$ is the scalar curvature computed using the Levi-Civita affine connection. Varying $S_{\rm GR_{(1)}}$ with respect to the metric yields the usual Einstein Field Equations for describing the interdependence between the derivatives of the metric field and the stress-energy of the matter content. However, an alternative way of arriving at the same set of equations is defining an action on metric affine spacetimes (which could, in principle, be geometrically richer than Riemannian spacetimes) but then using the method of Lagrange multipliers to constrain the variational calculus of extremizing the action to only torsion-free, metric-compatible manifolds. Then, the action for general relativity can be written as 
\begin{equation}
    S_{\mathrm{GR}_{(2)}} = \int \mathrm{d}^4 x \left[ \frac{\sqrt{-g}}{16\pi G} g^{\mu\nu} R_{\mu\nu}(\Gamma) + \lambda_{\alpha}^{\mu\nu} T^{\alpha}_{\mu\nu} + \hat{\lambda}^{\alpha}_{\mu\nu} Q_{\alpha}^{\mu\nu} \right].
    \label{eq:placeholder_label}
\end{equation}
where $\lambda$ and $\hat{\lambda}$ are Lagrange multipliers enforcing vanishing torsion $T$ and non-metricity $Q$, respectively\footnote{This means $\lambda$ and $\hat{\lambda}$ are treated as auxiliary fields such that variation of the action with respect to them imposes $T^{\alpha}{}_{\mu\nu}=0$ and $Q_{\alpha}{}^{\mu\nu}=0$. Because these constraints are integrable and holonomic, they can be solved and the resulting Levi-Civita connection substituted back into the action \citep{jimenez_geometrical_2019}.}. Note that the curvature $R_{\mu\nu}$ in the integral for $S_{\mathrm{GR}_{(2)}}$ is defined for a general affine connection $\Gamma$ (instead of $\mathcal{R}$ which uses Levi-Civita connection). 

As described earlier, TEGR and STEGR also possess only one non-vanishing geometric attribute (torsion and non-metricity respectively). Consequently, one can modify the form of $S_{\mathrm{GR}_{(2)}}$ by using a scalar corresponding to the other geometrical tensors defined in \S \ref{prelims} and modifying the Lagrange multipliers in order to constrain the rest so that they vanish. For teleparallel gravity, the action takes the form 
\begin{equation}
    S_{\mathbb{T}} = - \int \mathrm{d}^4 x \left[ \frac{1}{16 \pi G} \sqrt{-g}\,\mathbb{T}
    + \lambda_{\alpha}^{\ \beta \mu \nu} R^{\alpha}_{\ \beta \mu \nu}
    + \hat{\lambda}^{\alpha}_{\ \mu \nu} \nabla_{\alpha} g^{\mu \nu} \right] ,
    \label{eq:placeholder_label}
\end{equation}
where the Lagrange multipliers $\lambda$ and $\hat{\lambda}$ play the role of enforcing zero curvature and non-metricity respectively. The scalar $ \mathbb{T}$ is defined as
\begin{equation}
    \mathbb{T} \equiv -\frac{c_{1}}{4} T_{\alpha\mu\nu} T^{\alpha\mu\nu} - \frac{c_{2}}{2} T_{\alpha\mu\nu} T^{\mu\alpha\nu} + c_{3} T_{\alpha} T^{\alpha} ,
    \label{eq:placeholder_label}
\end{equation}
 with $T_\mu = T^\alpha_{\mu \alpha}$ being the trace of the torsion tensor and $c_1, c_2, c_3$ as free parameters.  \citet{jimenez_geometrical_2019} show that enforcing a vanishing curvature scalar $R=0$ and restricting the theory to metric-compatible connections yields 
\begin{equation}
    R = \mathcal{R}(g) + \mathring{\mathbb{T}} + 2\mathcal{D}_{\alpha} T^{\alpha} ,
    \label{eq:placeholder_label}
\end{equation} where $\mathring{\mathbb{T}}$ is just $\mathbb{T}$ after setting $c_1= c_2=c_3=1$. Notice, then, that the Ricci scalar defined for the Levi-Civita connection only differs from $\mathring{\mathbb{T}}$ by a sign and a total-divergence term, which integrates to a boundary term. Under the usual boundary conditions, $\mathcal{D}_{\alpha}T^\alpha$ therefore does not contribute to the bulk dynamics and can be omitted safely when deriving the field equations. In this way, the following action for a teleparallel theory of gravity
\begin{equation}
    S_{\mathrm{TEGR}} = - \int \mathrm{d}^4 x \left[ \frac{1}{16 \pi G} \sqrt{-g}\,\mathring{\mathbb{T} }(g, \Lambda)\right],
    \label{eq:TEGRAC}
\end{equation}
where a member $\Lambda$ of the general linear group $GL(4, \mathbb{R})$ is used to parametrize the affine connection $\Gamma$ in absence of curvature and non-metricity, is found to be dynamically equivalent to the Einstein-Hilbert action of GR. Here, I highlight a feature of $S_{\rm TEGR}$ that takes an important role later in my argument. The torsion scalar $\mathring{\mathbb{T}}$ depends independently on two degrees of freedom: first, the metric $g$ and, second, the affine connection $\Gamma$ (parametrized here by $\Lambda$). I will scrutinize this feature in greater detail later. 

The consequence and interpretation of this dynamical equivalence are discussed in the next subsection, but before that, I also consider the third node of the trio. While the case of STEGR is slightly more complicated, it follows the same general reasoning schema. Write an action over metric-affine spacetimes with a non-metricity scalar,
\begin{equation}
    S_{\mathbb{Q}} = - \int \mathrm{d}^4 x \left[ \frac{1}{16 \pi G} \sqrt{-g}\,\mathbb{Q} + \lambda_{\alpha}{}^{\beta \mu \nu} R^{\alpha}{}_{\beta \mu \nu} + \hat{\lambda}_{\alpha}{}^{\mu \nu} T^{\alpha}{}_{\mu \nu} \right],
    \label{eq:placeholder_label}
\end{equation}
where the non-metricity scalar is defined as
\begin{equation}
    \mathbb{Q} = \frac{c_{1}}{4} Q_{\alpha\beta\gamma} Q^{\alpha\beta\gamma}
- \frac{c_{2}}{2} Q_{\alpha\beta\gamma} Q^{\beta\alpha\gamma}
        - \frac{c_{3}}{4} Q_{\alpha} Q^{\alpha}
        + (c_{4} - 1) \tilde{Q}_{\alpha} \tilde{Q}^{\alpha}
        + \frac{c_{5}}{2} Q_{\alpha} \tilde{Q}^{\alpha}, 
    \label{eq:placeholder_label}
\end{equation} and  $Q_{\alpha} = Q_{\alpha \lambda}^{\lambda}$ and $\tilde{Q}_{\alpha} = Q^{\lambda}_{\lambda \alpha}$ are the two independent traces of the non-metricity tensor, $c_i$ corresponds to five free parameters and $\lambda$ represents the Lagrange multipliers tuned to enforce vanishing curvature and torsion.  Again, \cite{jimenez_geometrical_2019} show that for a torsion-free connection, the Ricci scalar can be related to the non-metricity scalar as
\begin{equation}
    R = \mathcal{R}(g) + \mathring{\mathbb{Q}} + D_{\alpha} \left( Q^{\alpha} - \tilde{Q}^{\alpha} \right),
    \label{eq:placeholder_label}
\end{equation}
where $\mathring{\mathbb{Q}}$ is just $\mathbb{Q}$ with $c_i=1$ for $i=\{1...5\}$. As before, in the case $R=0$, the total-divergence (i.e. the boundary) term $D_{\alpha} \left( Q^{\alpha} - \tilde{Q}^{\alpha} \right)$ can be dropped and the action is reduced to 
\begin{equation}
    S_{\mathrm{STEGR}} = -\frac{1}{16\pi G} \int \mathrm{d}^4 x \sqrt{-g}\,\mathring{\mathbb{Q}}(g,\xi),
    \label{eq:STEGRAc}
\end{equation}
which is dynamically equivalent to $S_{\mathrm{GR}_{(2)}}$ and corresponds to the theory named Symmetric Teleparallel Equivalent of General Relativity. Notice, again, that $\mathring{\mathbb{Q}}$ has two independent degrees of freedom i.e. the metric tensor $g$ and the affine connection $\Gamma$, which in this case (of vanishing torsion) is parametrized by a set of functions $\xi^\alpha$.  

 \subsection{Implications of Dynamical Equivalence}\label{implications}
Before turning to what dynamical equivalence implies about the physical content of these theories, it is worth making clear what it does \textit{not} imply. Actions that differ by boundary terms can yield theories that differ in several meaningful and non-trivial ways. For instance, it has been known for several decades that the Einstein-Hilbert action by itself does not lead to a well-posed variational problem on manifolds with a boundary. Thus, if one wants a well-posed Dirichlet problem in GR (black hole entropy calculations being a famous example), one has to include the Gibbons-Hawking-York term \citep{gibbons_action_1977, york_role_1972} which is constructed to exactly cancel the boundary contributions. Such additional terms are not required while varying $S_{\rm TEGR}$  or $S_{\rm STEGR}$ (because they only contain at most the first derivatives of the metric). More generally, the trinity theories can behave differently from each other whenever the asymptotic structure of spacetime is concerned. In general relativity, several important results are derived by defining quantities on a boundary at infinity after a conformal completion \citep{geroch_asymptotic_1977}, and such results are known to be very sensitive to geometrical properties (like smoothness and topology) of the boundary. For this reason, the physical content of general relativity that concerns asymptotic symmetry groups, the positivity of Arnowitt-Deser-Misner or Bondi mass, or the non-linear theory of gravitational radiation (used for generating template wave-forms for LIGO observations), would require explicit and focused analysis to show that they can be exported safely from GR to the other nodes of the Geometric Trinity. None of these results follow \textit{prima facie} from the dynamical equivalence described earlier (which is easily seen in that asymptotic analysis typically concerns quantities defined in covariant phase spaces and not the Einstein field equations directly, and no corresponding equivalence has been established at that level). 

Irrespective of these differences, \cite{wolf_underdetermination_2024} claim that the dynamical equivalence within Geometric Trinity theories is conceptually interesting for the following reason: if the actions that differ by a boundary term yield equivalent field equations, then the particle trajectories in all three theories coincide (barring a subtlety about the coupling of affine connection with matter field, which is where \S\ref{sec:chal} will locate the whole difficulty). Thus, insofar as probes of the spacetime geometry rely entirely on the paths of particles traced in spacetime, the Geometric Trinity theories remain indistinguishable. Note that dynamical equivalence implies more than mere empirical indistinguishability, which would be a claim contingent on current observational reach and could be broken in the future with access to better data. Instead, the stronger suggestion of the Geometric Trinity framework (which I attempt to deflate in the next section) is that no future refinement in precision measurements of particle trajectories would \textit{ever} be able to differentiate whether a given particle inhabits a solution of the GR field equations or their counterpart in TEGR or STEGR. While the assumption that the only probes of spacetime geometry are particle trajectories may seem questionable if unqualified (especially given my statements about gravitational radiation in the previous paragraph), for the purposes of this paper, I grant it \textit{arguendo}. The challenge developed in \S \ref{sec:chal} arises already at the level of particle trajectories, and needs no wider class of probes to show the incompatibility of the gauge gravity framework with Geometric Trinity's central proposal. 

I make one last comment in this subsection before turning to the gauge gravity framework, which stands in dialectical opposition to Geometric Trinity. There are various discussions in the philosophy of physics literature concerning whether the dynamical equivalence of Geometric Trinity leads to an under-determination in assessing statements about spacetime geometry\footnote{It is worth noting that GT proponents themselves acknowledge that the dynamical equivalence does not survive the passage to modified gravity theories. At the level of $f(Q)$, $f(R)$, and $f(T)$ extensions, the degeneracy is broken \citep{heisenberg_review_2024}. GT researchers suggest this as an empirical opening claiming that cosmological observations can constrain the choice between these modified theories. Although interesting both in their own right and in how they influence the understanding of underdetermination, these higher-order extensions of the gravitational Lagrangian do not have decisive observational support and do not directly concern the argument presented in this paper.} \citep{knox_newtoncartan_2011, mulder_is_2024, wolf_underdetermination_2024}. In particular, \citet{durr_invitation_2024} use this GT under-determination to constrain what ontological commitments remain palatable (notable removals from the list being realism about spacetime geometry, realism about spacetime curvature, etc.) and, also, to reinforce support for Geometric Conventionalism  (i.e. the view, in the tradition of Poincaré and Reichenbach, that the choice between empirically equivalent geometric descriptions of spacetime is conventional rather than factual). It is worth clarifying that in articulating the tensions between Geometric Trinity and gauge gravity, I do not attempt to undermine geometric conventionalism (even though the particular strategy employed by \citet{durr_invitation_2024}, which runs through GT's central proposal, does get undermined if the argument below succeeds). More broadly, I take no position on physics-informed-metaphysics here\footnote{On the question of physics-informed-metaphysics, I cite the wonderfully lucid Howard Stein --- `To borrow from the ancient philosophical tradition, what I believe the history of science has shown is that on a certain very deep question Aristotle was entirely wrong, and Plato — at least on one reading, the one I prefer — remarkably right: namely, our science comes closest to comprehending “the real”, not in its account of “substances” and their kinds, but in its account of the “Forms” which phenomena “imitate” (for “Forms” read “theoretical structures”, for “imitate”, “are represented by”).' \citep{stein_yes_1989}
} and, instead, frame this conflict as a philosophically interesting case study of a more general trade-off between epistemic virtues. Before doing that, though, I motivate and summarize some relevant aspects of the second framework of interest here i.e. Metric Affine Gauge Gravity.

\section{Metric Affine Gauge Theory of Gravity}
There is a long and rich history of trying to bridge the study of gauge symmetries with the study of spacetime geometry going back to the pioneering work by \citet{weyl_gravitation_1918,  weyl_gravitation_1929, utiyama_invariant_1956, kibble_lorentz_1961, sciama_analogy_1962}. The brief exposition here aggressively compresses it to a minimal conceptual core required for raising the `hypermomentum challenge' to the dynamical equivalence proposed by the Geometric Trinity. I begin by noting that, similar to how matter's motion is represented by an evolution of its external degrees of freedom, changes in the states of matter can be represented by an evolution of its internal degrees of freedom. Standard instances of such internal structure are dynamical quantities like charge, spin, and other properties that characterize \textit{what} matter is rather than \textit{where}. General relativity serves as a remarkable theory for understanding the evolution in external dimensions by describing it as a consequence of the non-linear relationship between spacetime geometry and its matter components. Correspondingly, the evolution of all internal degrees of freedom is well described by a family of gauge theories (as in the standard model of particle physics) that are built from considerations of local, non-rigid symmetry transformations of the target system. There is, then, a motivation to bring the two treatments together within a common framework.

Two possible strategies are available for bridging the mathematical understanding of evolution in internal and external dimensions. One may elevate the internal dimensions to the same status as the external ones by subjecting them to similar mathematical treatment (the most well-known consequence of which are the extra dimensions in Kaluza-Klein like theories \citep{kaluza_unification_1921, klein_quantentheorie_1926}). Alternatively, one may re-construct a theory resembling general relativity but via the procedures employed in the construction of gauge theories. The latter approach is what yields a gauge theory of gravity and, as I will show, it naturally motivates a generalization from using Riemannian spacetimes to using metric-affine spacetimes. 

A gauge theory of gravity can be assembled by following the aforementioned gauge procedure to extract dynamical laws from a given symmetry group. The first step is to identify the symmetry group of general relativity to feed into the gauge procedure. The symmetry group commonly associated with the arbitrarily curved spacetimes of general relativity is the diffeomorphism group $\rm Diff(M)$. However, this choice is unsuitable for accomplishing the goal of constructing a gauge theory of gravity. Unlike the groups corresponding to the gauge theories in particle physics, the diffeomorphism group is infinite dimensional and lacks enough structure to identify finite-dimensional representations suitable for defining matter fields \citep{hehl_metric-affine_1995, hehl_gauge_2014}. Elements of $\rm Diff(M)$ are not vertical automorphisms of any principal fiber bundle \citep{trautman_fiber_1980, gomes_same-diff_2022}, and they act also on the base manifold $M$ (instead of only affecting the fibers of a principal bundle $P \rightarrow M$ defined over the fixed base manifold $M$). To gauge gravity, then, one must begin with a different symmetry group. 

The Poincaré symmetry group $\text{ISO}(1,3) = \mathbb{R}^{1,3} \rtimes SO(1,3)$, composed of spacetime translations $P^a$ and Lorentz transformations $M^{ab}$, is the next natural candidate appropriate for producing a gauge theory of gravity \citep{kibble_lorentz_1961, sciama_analogy_1962}, since this is the global symmetry group associated with special relativity. The physical motivation comes from Noether's first theorem that demonstrates invariance under spacetime translations and Lorentz transformations generates, respectively, the conserved stress-energy tensor $T^{\mu\nu}$ and the conserved angular momentum tensor with an intrinsic spin current $S^{\mu\nu\rho}$ \citep{hehl_general_1976}. These are precisely the currents that characterize the matter content of any relativistic field theory and, thus, the symmetry group whose conserved charges are carried by matter becomes a natural candidate for constructing a theory of gravity. In what follows, I build the simplest gauge theory of gravity arising from the Poincaré group (namely, Einstein-Cartan theory) and show that it already contains the essential challenge to the Geometric Trinity. Later, I describe how the argument extends to the full metric-affine setting and puts pressure on all three nodes of the trinity. 

The gauge procedure associates a potential to each generator of the symmetry group. Corresponding to the translation generators $P^a$, it yields a set of one-form fields $\vartheta^a = e^a_{\ \mu} dx^\mu$, called the co-frame (or vierbein). Corresponding to the Lorentz generators $M^{ab}$, it yields the connection one-forms $\Gamma^a_{\ b} = \Gamma^a_{\ b\mu} dx^\mu$, called the affinity. These two objects are the fundamental dynamical variables of the theory, playing the role that the metric $g_{\mu\nu}$ and the connection coefficients $\Gamma^\alpha_{\mu\nu}$ play in the standard formulation of GR. To make contact with the coordinate-index notation used in the preceding sections, the spacetime metric is recovered from the co-frame according to
\[
g_{\mu\nu}=\eta_{ab}\,e^a{}_\mu e^b{}_\nu,
\]
where \(\eta_{ab}\) is the Minkowski metric on the local frame. The linear connection one-form \(\Gamma^a{}_{b\mu}\) is related to the coordinate affine connection \(\Gamma^\rho{}_{\nu\mu}\) through the co-frame (and its derivatives). Thus, once a co-frame is specified, the frame-index and coordinate-index formulations provide equivalent descriptions of the affine connection.

From the gauge potentials  (i.e. the co-frame and the affinity), one constructs the field strengths by taking their exterior derivatives\footnote{Exterior derivatives $d$ are simply the generalization of derivative operator to one-forms, such that $dA = \partial_\mu A_\nu dx^\mu \wedge dx^\nu$.} and adding the appropriate wedge product\footnote{A wedge product is an anti-symmetric map that takes two differential forms and produces a higher-order form  $dx^i \wedge dx^j = -dx^j \wedge dx^i$ .} terms to account for the non-commutativity of Poincaré group transformations. Corresponding to the co-frame and affinity respectively, the field strengths are the torsion and the curvature 
\begin{equation}
T^{\alpha} \equiv d\vartheta^{\alpha} + \Gamma^{\alpha}_{\ \beta} 
\wedge \vartheta^{\beta} = \frac{1}{2} T^{\alpha}_{\ ij}\, dx^{i} \wedge dx^{j}, 
\qquad
R^{\ \beta}_{\alpha} \equiv d\Gamma^{\ \beta}_{\alpha} - \Gamma^{\ \gamma}_{\alpha} 
\wedge \Gamma^{\ \beta}_{\gamma} = \frac{1}{2} R^{\ \beta}_{ij\alpha}\, 
dx^{i} \wedge dx^{j},
\label{eq:field_strengths}
\end{equation}
which the reader will recognize as the torsion tensor $T^\alpha_{\ \mu\nu}$ and Riemann tensor $R^\alpha_{\ \beta\mu\nu}$ from \S\ref{prelims}, now expressed as differential forms. Having identified certain 2-forms as the field strengths of the gauge potentials, I proceed to the final step of assembling and varying an action for the dynamics of the field strength. The geometrical constraint on the action is that it needs to be a 4-form, so that it can be integrated over a 4D manifold. One possibility then is to construct a 4-form by wedging the field strength 2-form $R_{\alpha\beta}$ with the 2-form $\eta_{\alpha\beta}\equiv\frac{1}{2}\epsilon_{\alpha\beta\gamma\delta}\,\vartheta^{\gamma} \wedge\vartheta^{\delta}$ (i.e. the dual of the simplest 2-form constructed by wedging two co-frames)\footnote{Note that $\eta_{\alpha\beta}$ here written with Greek indices is distinct from the Minkowski metric $\eta_{ab}$ written with Roman ones.}. Then, ignoring the cosmological constant term, the action assembled becomes 
\begin{equation}
S_{\rm EC} = \frac{1}{2\kappa}\int_M \left[ R^{\alpha\beta} \wedge \eta_{\alpha\beta}  \right] + S_m
\label{eq:EC_action}
\end{equation}
where, unlike in the preceding discussion of the actions of TEGR and STEGR, the matter action term $S_m$ is now explicitly included in the expression. The EC in the subscript of Eq. \eqref{eq:EC_action} refers to the Einstein-Cartan action (or Einstein-Cartan-Sciama-Kibble action). Generically, like the dynamical degrees of freedom of the geometric part of the action, the matter Lagrangian $\mathcal{L}_m$ can depend on both gauge potentials, a dependence that becomes crucial to the disagreement with the Geometric Trinity, such that 
\begin{equation}
    S_m = S_m (\vartheta, \Gamma) \equiv \int \mathcal{L}_m\left(\vartheta, \Psi, D\Psi\right)
\end{equation}
where $\Psi$ is the matter field and the affinity couples to matter action through the covariant derivative of the matter field $D\Psi = d\Psi + \Gamma\Psi$. Notice, until this point, the expression above for Einstein-Cartan action is structurally similar to that of Einstein-Hilbert action and differs only in the explicit inclusion of a matter term $S_m$ depending on two dynamical variables (reflecting that it inhabits not a Riemannian spacetime but a generic Riemann-Cartan spacetime possessing both curvature and torsion).

I now address the issue of two dynamical degrees of freedom which I flagged earlier. Within the gauge gravity literature (and wider physics in general), unless discussed explicitly, one derives the field equations by varying the action with respect to all independent dynamical variables to arrive at the complete set of equations describing the field dynamics. Varying the action with respect to the co-frame gives the set of equations \citep{hehl_gauge_2014}, 
\begin{equation}
    \mathrm{Ric}_{\alpha}{}^{i} - \frac{1}{2} e_{\alpha}^{i} \mathrm{Ric}_{\gamma}{}^{\gamma} = \frac{\kappa}{e} \mathcal{T}_{\alpha}{}^{i} \, , \label{eq:placeholder_label}
\end{equation}
that relate the Ricci curvature tensor computed for the full Riemann-Cartan connection $\rm Ric \left(\vartheta, \Gamma\right)$ to the canonical energy-momentum tensor of matter $\mathcal{T}$ (serving as the analog of Einstein-Field equations for Riemann-Cartan geometry, expressed in terms of the vierbein formalism). These equations, however, are not a complete description of the system's dynamics because they do not include the result of varying the action with respect to affinity. Doing so yields the following equations \citep{hehl_gauge_2014}
\begin{equation}
    \mathrm{Tor}_{\alpha \beta}{}^{i}
    - e_{\alpha}^{i} \, \mathrm{Tor}_{\beta \gamma}{}^{\gamma}
+ e_{\beta}^{i} \, \mathrm{Tor}_{\alpha \gamma}{}^{\gamma}
    = \frac{\kappa}{e} \,\mathscr{S}_{\alpha \beta}{}^{i},
    \label{eq:placeholder_label}
\end{equation}
that relate the torsion of spacetime $\rm Tor$ with the spin angular momentum density of matter $\mathscr{S}_{\alpha\beta}{}^{i} \equiv \delta \mathcal{L}_m /\delta \, \Gamma^{\alpha\beta}{}_{i}$\footnote{Starting with a slightly more general form of the EC action that includes terms quadratic in torsion and curvature yields the Poincaré Gauge Theory (PGT) \citep{hehl_general_1976, weatherall_are_2025}, which is also a curvature- and torsion-full description of gravity but with the additional feature of admitting torsion propagation in vacuum spacetimes (similar to how GR admits curvature propagation outside of matter).}. For a scalar field the spin angular momentum density always vanishes. More strikingly, the Maxwell field contributes nothing either because, although it carries helicity, it has no gauge-covariant spin and remains insensitive to torsion as well \citep{hehl_gauge_2014}. But the spin angular momentum density can be non-zero in the case of Dirac spinors (which are useful to model particles like electrons and quarks but their relationship with spacetime torsion is yet to be experimentally established). This second set of field equations, concerning the affine connection and the intrinsic spin of matter, is not considered in standard Geometric Trinity equivalence arguments.\footnote{Readers familiar with the Palatini formulation of GR will recognize the independent variation of metric and connection. Palatini GR does include a connection equation, but with a matter action independent of the connection, it recovers the Levi--Civita connection up to projective freedom.} Its omission is the source of the incompatibility between gauge gravity and the Geometric Trinity. I now explore the nature and implications of this tension, extend this comparison to a more general case (i.e. Metric Affine Gravity) and evaluate what epistemic warrants one can appeal to in defense of these two frameworks.

\subsection{The Upshot and The Challenge}\label{sec:chal}
The two sets of field equations derived above admit a compact summary: in the Riemann-Cartan formulation of gravity, energy-momentum sources curvature and spin angular momentum sources torsion. The first relationship is the familiar one from GR; the second is an additional dynamical ingredient. It implies that matter with non-trivial internal structure (specifically, matter carrying a non-vanishing spin angular momentum density $\mathscr{S}_{\alpha\beta}{}^{i}$) couples differently to spacetime geometry than structureless matter. In a torsionful spacetime, the trajectories of particles with intrinsic spin deviate from those in a torsion-free spacetime, even when the curvature is identical or vanishing \citep{iosifidis_motion_2024}.

Juxtaposing this with the Geometric Trinity claims makes the tension manifest. GT asserts an underdetermination of spacetime geometry on the basis that particle trajectories are degenerate across corresponding configurations of curvature, torsion, and non-metricity, rendering the three nodes of the trinity in-principle indistinguishable. The gauge gravity framework rejects this indistinguishability because for matter with non-vanishing spin angular momentum, there is no dynamical equivalence on which the underdetermination can be grounded. According to the gauge gravity framework, GT's equivalence claim implicitly assumes $\mathscr{S}_{\alpha\beta}{}^{i} = 0$, an assumption MAG holds to be unmotivated within the gauge-theoretic framing. In \S \ref{sec:epv}, I return to discuss how a Geometric Trinity proponent could respond to this charge; for now, let us develop this challenge further.

Granting validity to the gauge procedure, thus, deflates the GT framework's claim of in-principle indistinguishability to one of contingent empirical indistinguishability. The prediction of gauge gravity is that spacetime possesses torsion where matter possesses intrinsic spin. However, detection of this spacetime torsion remains challenging. Unlike mass-energy, intrinsic spin tends to cancel when aggregated into macroscopic bodies (and thus observations from Gravity Probe B as discussed by \citet{mao_constraining_2007} do not really track the spin-torsion coupling relevant to the present discussion). Yet, proposals exist for detecting torsion signatures \citep{hehl_poincare_2013, puetzfeld_prospects_2014} using both table-top experiments and astrophysical observations and, while the experimental program remains nascent, there are already dedicated conferences for the field\footnote{See: TORSION 2026 — Tests Of Relativity with Spin InteractiONs, \url{www.nucleares.unam.mx/torsion}. The deflation of the claim from dynamical to empirical equivalence bears on assessing the pursuit-worthiness of this field. For instance, someone convinced of an under-determination between curved and twisted spacetimes may not find pulsar-based searches for spacetime torsion productive telescope time. On the other hand, those skeptical of the GT may wish to proceed to build such experiments regardless.}.   

The challenge extends as one generalizes the gauge group. Replacing the Poincaré group $\text{ISO}(1,3)$ with the full affine group $GA(4,\mathbb{R}) = \mathbb{R}^4 
\rtimes GL(4,\mathbb{R})$ liberates the connection from the metric-compatibility condition, allowing non-metricity alongside torsion and curvature in the manifold. The sourcing current in this more general theory, the metric affine theory of gravity, generalizes to the full hypermomentum tensor $\Delta_{\lambda}{}^{\mu\nu} \equiv -\frac{2}{\sqrt{-g}}\frac{\delta S_m}{\delta \Gamma^{\lambda}{}_{\mu\nu}}$, decomposing into spin, dilation, and shear under $GL(4,\mathbb{R})$ \citep{hehl_metric-affine_1995}. Spin-angular momentum, discussed above in the context of Einstein-Cartan, corresponds (up to normalization conventions) to the anti-symmetric part of the more general object that is the hypermomentum tensor. Now, STEGR also loses its equivalence claim because matter with non-vanishing intrinsic dilation or shear would source spacetime non-metricity in the gauge gravity framework. The GT under-determination, already challenged at the Einstein-Cartan level, is fully undermined in the metric-affine setting if hypermomentum is not assumed to be vanishing.

\subsection{On Hypermomentum}
Clearly, the disagreement as stated above depends sensitively on how one treats the hypermomentum tensor (hence, the choice of the name `hypermomentum challenge'). A Geometric Trinity proponent can still remain skeptical of the physical significance one ought to afford to, what they may regard as, a new object introduced to break the dynamical equivalence. In \S \ref{sec:epv}, I show that evaluating the epistemic virtues of GT and MAG is essentially an exercise in contrasting different attitudes towards the hypermomentum tensor. Prior to that, however, I assemble the material such an assessment requires (where the object comes from and what is known about its capacity to generate observables).

The energy-momentum tensor $T_{\mu\nu}$ has received sustained philosophical attention in the past. Considerations about energy conditions, localizability and coupling-procedures have generated substantial discourse amongst physicists and philosophers (for examples, see \citet{hoefer_energy_2000, lehmkuhl_mass-energy-momentum_2011,curiel_primer_2014, ferreiro_deflating_2025}). Hypermomentum has not been subjected to analogous foundational scrutiny and it remains unclear what aspects of the accounts developed in context of energy-momentum (e.g. Curiel's impossibility argument against a gravitational $T_{\mu\nu}$ being a concomitant of the metric in the language of jet-bundles \citep{curiel_tensorial_2012})  map onto the hypermomentum tensor $\Delta^\lambda_{\mu\nu}$. Nevertheless, investigations of hypermomentum have carved out a niche within the gauge gravity community and, here, I mention some results relevant as a foothold for future philosophical inquiry.

One reason hypermomentum may have received comparatively little attention in philosophical discussions of spacetime geometry is the strong and specific interpretive picture with which it has often been associated. A heuristic route by which hypermomentum has been physically motivated in the gauge-gravity literature comes from an analogy, due especially to Hehl and collaborators \citep{hehl_hypermomentum_1976, hehl_hypermomentum_1976_2}, with a generalization of continuum mechanics. Historically, the analogy goes back to the Cosserat brothers' work on generalized elasticity \citep{cosserat_theorie_1909} that Cartan explicitly took as inspiration for his work on torsion a decade later \citep{cartan_sur_1922, shapiro_physical_2002}. In the theory of polar continuum media, one extends ordinary elasticity theory by allowing the microscopic constituents of a medium to possess internal degrees of freedom, including independent rotations and more general deformations. The associated response currents decompose into parts corresponding, again, to spin, dilation, and shear. The literature surrounding Einstein--Cartan theory, and later metric-affine gravity, takes this analogy between microstructure dynamics of polar continuum media and internal degrees of freedom of spacetime very seriously. For instance, in a section titled `Spacetime as a 4-dimensional Elastic Continuum', \citet{hehl_hypermomentum_1976} claim, `[The mathematical continuum representing spacetime], in the light of Einstein's theory of gravitation, the general theory of relativity (GR), is neither static nor rigid, but reacts elastically to its matter distribution.' Such a treatment of spacetime as an elastic substance may invite skepticism, not only from relationalists, but from anyone reluctant to attribute microstructure to spacetime itself. 

The challenge raised against the dynamical equivalence claims of GT, however, is insensitive to whether one is willing to admit a treatment of spacetime as an elastic medium with microstructure. I emphasize that the polar-continua analogy is only one potential interpretation of hypermomentum and my observations regarding the tension between GT and MAG are agnostic towards the chosen interpretation. For the purposes of this paper, one may even adopt principled silence over questions of ontology or interpretation, and simply treat the tensor as the object that is naturally obtained when applying the gauge procedure to the general affine group (similar to how spin angular momentum naturally comes from constructing a gauge theory using the Poincar\'e group). In this way one can understand the disagreement between the two frameworks on the basis of features completely internal to them, without getting caught in further questions concerning interpretations. The mathematical status of hypermomentum within metric-affine or Poincar\'e-gauge frameworks is perfectly unambiguous: it is the source conjugate to the affine connection. In fact, the mathematical and conceptual problems flagged at the beginning of \S\ref{sec:GT} in connection with teleparallel and symmetric teleparallel approaches (surplus structure and mathematical coherence) have no comparable counterparts for Poincar\'e gauge gravity and metric-affine gauge gravity. The latter theories, thus, appear to be comparatively more robust\footnote{In context of comparing the gauge-status of tele-parallel theories and Riemann--Cartan theory,  Jim Weatherall says, ``Geometrically, the Riemann-Cartan geometry on which PGT is based is unimpeachable, and I am not aware of any criticisms that theory is mathematically problematic'' \citep{weatherall__2025}. }. 

Another aspect of hypermomentum that bears on whether it can make observational contact, besides the ones gestured at in \S\ref{sec:chal}, has been developed into a study of phenomena involving \emph{hyperfluids} \citep{obukhov_hyperfluid_1993, iosifidis_perfect_2021}. Beginning with work by Obukhov and Tresguerres and subsequently extended in later treatments, there has been an interest in developing cosmological solutions that feature non-vanishing hypermomentum and are governed by the dynamics of metric-affine gravity \citep{iosifidis_cosmological_2021}. Within Friedmann--Lema\^{\i}tre--Robertson--Walker symmetry, such models source specific combinations of torsion and non-metricity and thereby modify the effective cosmological dynamics. More recent work has derived generalized Friedmann equations for quadratic metric-affine theories in the presence of cosmological hyperfluids \citep{iosifidis_cosmology_2023} and begun attempts at making contact with observational data \citep{chaudhary_yano_2025}. What remains much less developed, at least relative to the mature cosmological phenomenology of some other modified-gravity frameworks, is a systematic bridge from these homogeneous solutions to nonlinear structure formation and other precision probes of late-time cosmology. I mention this to illustrate that, even though many open questions remain about the empirical accessibility of hypermomentum, it is still a sufficiently well-defined quantity that can facilitate potential observables in the future. 

\section{Assessing Epistemic Virtues}\label{sec:epv}
Having discussed the technical details that generate the tension between the Geometric Trinity and gauge gravity frameworks, I highlight the more general epistemological lessons that this case study represents. To do so, I must anticipate the objections and counter-objections that Geometric Trinity proponents could raise against the hypermomentum related challenge.

I begin by restating the tension. GT claims that certain spacetime theories of gravity based on torsion, curvature and non-metricity respectively are dynamically equivalent (hence, in-principle indistinguishable) whereas no such equivalence exists generically within the gauge gravity framework (which holds the different geometric attributes to be in-principle distinguishable). I tracked this tension to the differing attitudes towards assumptions on vanishing or non-vanishing of hypermomentum.

Given what I said about the observational evidence of hypermomentum, the most natural defense for GT could be an appeal to parsimony and a charge against metric affine gravity of speculatively proliferating new currents from a richer matter-sector to break the equivalence. In a stronger version of this defense, one can claim that hypermomentum does not even exist as an object in the GT theories by virtue of the Lagrange multiplier choices. Hence, its vanishing is not to be considered a matter-sector assumption at all. In other words, it is meaningless to accuse the framework of positing a quantity as vanishing if the corresponding object is not even defined within the theory. If this defense of dynamical equivalence passes, then the burden of proof effectively shifts to the gauge gravity framework to demonstrate that intrinsic spin does indeed couple to torsion (which, as discussed in the previous sections, is an open and difficult challenge). There are certain problems, however, that resist this shifting in the burden of proof. 

The first objection against the parsimony of the trinity theories is to assert that the hypermomentum object \textit{does} already exist within the GT framework, even if it is never explicitly defined or discussed. This is because, following standard practice, one has to vary the action with all degrees-of-freedom. Even within the GT framework, the actions have a dependence on a second independent dynamical variable (which is what I flagged for $\mathring{\mathbb{T} }(g, \Lambda)$ and $\mathring{\mathbb{Q}}(g,\xi)$ in \S\ref{sec:GT_Arg}). Variation of the action with the connection (i.e. with $\Lambda$ and $\xi$ in Eq. \eqref{eq:TEGRAC} and Eq. \eqref{eq:STEGRAc} respectively) would generate the corresponding hypermomentum-like object within the GT framework and supply the second set of field equations for the theory. More strongly, to assume one entire sector of the dynamical structure (i.e. second set of field equations) as trivial is a speculative excess that is neither derived from within the theory nor explicitly justified on the basis of other principles. The Lagrange multipliers are only set-up over the geometric sector and how they bear on the matter Lagrangian is left completely undiscussed in establishing the argument for dynamical equivalence. 

One could respond that this is still a non-issue for GT, since the standard practice (of varying with all dynamical variables) binds your theory-building, in this context, more strongly if your aim is to construct a gauge theory of gravity. Instead, if the aim is explicitly to establish the dynamical equivalence between the trinity theories, one is not beholden to the standard implementation of the gauge procedure\footnote{An important issue arises with this retreat. \citet{jimenez_geometrical_2019} explicitly take TGR as a gauge theory and, hence, cannot discard the gauge approach \textit{a priori}. Although, as mentioned earlier, there are strong unresolved objections against the gauge status of TGR or STGR.}. This retreat is available, but it requires substantially modifying the central proposal of GT because they must grant that the equivalence is no longer between the trinity theories simpliciter but only under additional \textit{empirically contingent} matter sector assumptions, thereby deflating the claim of \textit{in-principle} indistinguishability. In light of this, gauge gravity theorists can push-back against the speculative excess charge raised against them earlier by claiming a different kind of parsimony.

It is worth isolating explicitly three distinct (albeit inter-related) forms of parsimony --- ontological parsimony (the virtue of positing fewer fundamental constituents in physical theories), epistemic parsimony (the virtue of positing fewer assumptions lacking epistemic warrants) and methodological parsimony (the virtue of positing fewer physical `sectors' requiring independent theory-building procedures). Geometric Trinity's dismissal of hypermomentum was a claim to ontological parsimony, insofar as it could be established that hypermomentum plays no role in understanding current observations and that hypermomentum object simply does not exist within the GT framework. If, as I have argued, the object does exist within the GT framework and must instead be set to zero by stipulation, gauge gravity theorists can object that this stipulation is insufficiently justified and claim an epistemic parsimony for declining to make it.
GT may nonetheless press their original claim that, absent any evidence for believing in spin-torsion coupling, it holds an ontological parsimony by not populating the theory with additional entities that serve no explanatory role. In response, gauge-gravity theorists press a further claim of a methodological parsimony. Implementing the gauge principle in constructing both particle physics theories \textit{and} gravitational physics theories grounds them both in the same underlying procedure and, thus, enhances the inter-theoretic coherence between the two sectors. 

There are various instances of research programs that trade one flavor of parsimony to gain another. In one reading, the support for GR itself can be claimed as a kind of ontological inflation (positing curvature as a new property of spacetime) in order to implement the equivalence and relativity principles (which famously deflate the assumption that there exists a class of privileged observers). This can be thought of as historical precedence for mathematical sophistication or ontological inflation as a cost of implementing a principle that achieves epistemic or methodological deflation. An ontological inflation, thus, is not \textit{prima facie} problematic but, rather, the introduction of the novel entities must be evaluated with a wider perspective on what they enable for  the theory\footnote{Kuhn identifies several epistemic virtues commonly invoked in discussions concerning theory choice. Beyond accuracy and consistency, these include scope, simplicity and fruitfulness \citep{kuhn_objectivity_1977}. In the present case, metric-affine gauge gravity gains in scope and potential fruitfulness by accommodating matter microstructure and possible new empirical contacts, whereas the Geometric Trinity retains an advantage in mathematical simplicity. Kuhn nevertheless argues that such criteria do not uniquely determine theory choice, since they may be interpreted differently and may pull in competing directions \citep{kuhn_objectivity_1977}.} (just as we have done in the preceding sections). 

Still, as a final line of defense, one could be a skeptic against the fundamental desideratum of Riemann--Cartan and Metric--Affine Gravity to cast GR as a Yang-Mills like gauge theory. One might point to fundamental differences between internal and external degrees-of-freedom (and their corresponding mathematical structures) to highlight why gauging gravity is a problematic endeavor. There are replies available from both philosophers and physicists. For instance, \citet{wallace_fields_2015} argues that on a parametrized representation of field theory, internal and spacetime symmetries can be treated on par. A more detailed presentation of the motivations and merits of gauging gravity and extending it to the affine group is given in \citep{hehl_metric-affine_1995, hehl_gauge_2014}, where it is highlighted that as long as a conserved current and a corresponding symmetry group are in hand, the gauge procedure is indeed applicable. An even stronger skeptic may raise concerns about utilizing the gauge procedure as a legitimate theory building device, irrespective of its past successes, citing issues with over-reliance on symmetry arguments given the absence of an account of why they should work in the first place. Even this stronger worry has been engaged with directly. \citet{gomes_gauge_2025}, for instance, reconstructs the gauge procedure as a `Noether gauge argument' which aims to supply a principled rationale for the procedure's fruitfulness rather than a bare appeal to its past successes. However, summarizing and evaluating the status of these extensive and more general debates falls beyond the scope of this paper, whose sharp focus is on the hypermomentum challenge to GT in particular.

A strong reading of the account presented in this section could be that if one is generally sympathetic to gauge theories and finds the demand to gauge gravity acceptable, then the claims of equivalence by Geometric Trinity researchers are unconvincing because they depend on matter sector assumptions that are left implicit and appear ad-hoc. The more modest reading of the account here, which I feel more comfortable endorsing, is that -- 1) a practicing physicist makes trade-offs between various epistemic virtues in evaluating the pursuit-worthiness of a framework that are better left acknowledged than implicit or unexamined; and 2) the incompatibility between the Geometric Trinity and gauge gravity frameworks is one instantiation of this general trade-off. 

\section{Conclusion}
The comparison developed in this paper reveals a layered form of underdetermination. At the first level, the Geometric Trinity asserts an underdetermination between spacetime curvature, torsion, and non-metricity by an appeal to dynamical equivalence between GR, TEGR and STEGR. At a second level, this underdetermination is challenged by following the gauge-procedure towards a theory of gravity, which yields matter couplings (of spin and, more generally, hypermomentum) for which no such dynamical equivalence obtains. Given the unclear observational status of hypermomentum, this results in a meta-underdetermination: a disagreement over whether the original underdetermination is physically salient or an artifact of restrictive assumptions on the matter sector. I argue that this is the philosophically interesting locus of the MAG-GT conflict. While GT claims that the different geometrical objects can be reduced into representations of the same real world phenomena, the gauge gravity framework considers it more fruitful to associate different features of the world with distinct mathematical objects. Thus, it is a disagreement about where to place the speculative burden between the framework that introduces hypermomentum as a novel physical current and the framework that sets it to zero. How that burden should be distributed, however, turns on broader questions concerning empirical access, forms of parsimony in the absence of discriminating evidence, and competing desiderata of theory construction.

I have not claimed that the gauge procedure uniquely mandates any particular interpretation of hypermomentum, nor have I taken a position on geometric conventionalism broadly construed or on spacetime ontology. What the argument does establish is that the dynamical equivalence invoked by the Geometric Trinity is conditional on substantive assumptions about matter coupling that are rejected within metric-affine gauge gravity. The resulting conflict is therefore not merely interpretive; it concerns which dynamical degrees of freedom and source currents a viable theory of gravity ought to admit.

The disagreement therefore has an empirical bearing, even if meaningful probes of hypermomentum remain beyond current capabilities. Evidence for non-trivial hypermomentum couplings would break the claimed in-principle indistinguishability of the trinity theories and vindicate the gauge-gravity objection to unrestricted dynamical equivalence. Continued empirical inaccessibility would not establish the Geometric Trinity, but would strengthen its appeal to ontological and theoretical parsimony. The central contribution of this paper is to assess this fault line and to make the nature and stakes of the associated choices clearer.

\section*{Acknowledgments}
I am grateful to Friedrich Hehl and Erik Curiel for discussions that seeded this line of inquiry and Dennis Lehmkuhl, James Read and Gal Ben-Porath for extensive feedback on the drafts. I also thank Henrique Gomes and various attendees of UPAC `Spacetime Matters' conference for helpful suggestions. 

\bibliography{MAG-GT}
\bibliographystyle{plainnat}

\end{document}